\documentclass{llncs}

\usepackage[utf8]{inputenc} 

\usepackage[hyphens]{url}  
\usepackage{graphicx} 
\usepackage{algorithm}
\usepackage{multirow}
\usepackage{makecell}
\usepackage{xspace}
\usepackage{subcaption}
\usepackage{array}
\usepackage{tabularx}
\usepackage{array}
\usepackage{hyperref}

\usepackage{textcase}
\usepackage[noabbrev, capitalise]{cleveref}
\usepackage{newfloat}
\usepackage{listings}
\floatstyle{ruled}
\newfloat{listing}{tb}{lst}{}
\floatname{listing}{Listing}
\usepackage{xcolor}

\usepackage{xcolor}
\usepackage{soul}

\usepackage{algpseudocode} 

\begin{document}

%
\title{Temporal Motifs for Financial Networks: A Study on Mercari, JPMC, and Venmo Platforms}

\titlerunning{Temporal Motifs for Financial Networks}  
%
\author{Penghang Liu\inst{1} \and
    Bahadir Altun\inst{2} \and
    Rupam Acharyya\inst{3}\thanks{This work was done when the author was with University at Buffalo} \and
    Robert E. Tillman\inst{4}\thanks{This work was done when the author was with J.P. Morgan Chase AI} \and
    Shunya Kimura\inst{5} \and
    Naoki Masuda\inst{6,}~\inst{7} \and
    Ahmet Erdem Sar{\i}y\"{u}ce \inst{2}
}
\authorrunning{Liu et al.} 

\institute{
J.P. Morgan Chase AI \and
Department of Computer Science and Engineering, University at Buffalo \and
Amazon \and
Optum Labs \and
Mercari, Inc. \and
Department of Mathematics and Computational, University at Buffalo \and
Data-Enabled Science and Engineering Program, University at Buffalo \\
Contact person: \email{erdem@buffalo.edu}
}

\maketitle

\begin{abstract}
Understanding the dynamics of financial transactions among people is critical for various applications such as fraud detection.
One important aspect of financial transaction networks is temporality. The order and repetition of transactions can offer new insights when considered within the graph structure.
Temporal motifs, defined as a set of nodes that interact with each other in a short time period, are a promising tool in this context.
In this work, we study three unique temporal financial networks: transactions in Mercari, an online marketplace, payments in a synthetic network generated by J.P. Morgan Chase, and payments and friendships among Venmo users.
We consider the fraud detection problem on the Mercari and J.P. Morgan Chase networks, for which the ground truth is available.
We show that temporal motifs offer superior performance to several baselines, including a previous method that considers simple graph features and two node embedding techniques (LINE and node2vec), while being practical in terms of runtime performance.
For the Venmo network, we investigate the interplay between financial and social relations on three tasks: friendship prediction, vendor identification, and analysis of temporal cycles.
For friendship prediction, temporal motifs yield better results than general heuristics, such as Jaccard and Adamic-Adar measures.
We are also able to identify vendors with high accuracy and observe interesting patterns in rare motifs, such as temporal cycles.
We believe that the analysis, datasets, and lessons from this work will be beneficial for future research on financial transaction networks.
\end{abstract}

\section{Introduction}
Financial relationships become increasingly important in today's society. 
The quick development of online payment services, such as Apple Pay, PayPal, Stripe, and Venmo, has revolutionized people's daily financial activities.
Online payment services have become the most widely used payment method and generate a tremendous amount of transactions every day.
Critical applications, such as fraud detection, anti-money laundering, and link prediction, require a thorough understanding of the dynamics in financial transaction networks in various media.

One key characteristic of financial transactions is temporality.
The order of transactions, time differences, and the way those appear within the graph structure can offer crucial insights.
Temporal motifs, defined as a set of nodes that interact with each other in a short time period, are shown to be a powerful tool in this context~\cite{liu2021temporal,Liu22,K11,S14,H15,P17}.
Temporal motifs are used in various domains such as communication networks~\cite{zhao2010communication}, trading~\cite{bajardi2011dynamical}, human contact networks~\cite{zhang2015human}, and others~\cite{chechik2008activity,Faisal14,jurgens2012temporal,kovanen2013temporal,li2014statistically,xuan2015temporal}.

In this work, we study fraud detection, link prediction, and node classification problems by using temporal motifs on three novel temporal financial networks:
\begin{itemize}
\item {\bf Transactions in Mercari, an online marketplace.} We build the consumer-to-consumer online marketplace network of Mercari, which is one of the largest e-commerce platforms in Japan. We obtain the network through personal correspondence. An event $(u, v, t)$ denotes that user $u$ sells an item to user $v$ at time $t$. It is illegal to sell certain items, such as weapons, medicine, and used underwear, and the sellers of those items are marked as fraudulent users. 
\item {\bf Payments in a synthetic network generated by J.P. Morgan Chase (JPMC).} We consider a synthetic payment network generated by JPMC~\cite{assefa2020generating}, obtained via personal correspondence. The network preserves many unique features observed in the real transactions and has been used for JPMC internal research. Importantly, each transaction contains a monetary amount information and marked as licit or fraudulent. A user is a fraudster if s/he receives at least one fraud payment.
\item {\bf Payments and friendships among Venmo users.} We identify 600 most active users in a public dataset of seven million Venmo transactions~\cite{venmo7m} and collect the transaction history of these users using the Venmo API. We also collect the friendship relations among all the users. Each transaction comes with the IDs of sender and receiver, the timestamp, and a payment note.
\end{itemize}

We study fraud detection on the Mercari and JPMC networks using temporal motifs from one-hop egocentric networks. Each user is represented by a feature vector based on motif counts. Using logistic regression, SVM, random forest, and XGBoost, we show that temporal motif features outperform baselines---simple graph features~\cite{kodate2020detecting}, LINE~\cite{LINE}, and node2vec~\cite{node2vec}---reaching up to 0.89 AUC on Mercari and 0.82 on JPMC. On Venmo, we explore friendship prediction, vendor detection, and temporal cycles. Motif-based features outperform heuristics in predicting friendships. For vendor detection, we manually label users and release the dataset. We also identify rare cyclic patterns ($A \rightarrow B \rightarrow C \rightarrow A$) linked to behaviors like poker games. Our results and datasets offer valuable insights for future research.

Our contributions can be summarized as follows:
\begin{itemize}
\item We study three novel financial transaction networks, Mercari, JPMC, and Venmo, which are a diverse set of media for financial activities.
\item We utilize temporal motifs for fraud detection on the Mercari and JPMC networks where the fraud activity is defined differently.
\item We investigate the interplay between financial and social relationships on the Venmo network by using temporal motifs. We consider friendship prediction and vendor identification problems, and also analyze temporal cycles.
\item We release financial transactions, friendships, and vendor labels for Venmo network, and also share the codes at \url{https://github.com/erdemUB/ASONAM25}.
\end{itemize}

\noindent {\bf Remark.} We conduct experiments on three datasets: Mercari, JPMC, and Venmo. These data are acquired from real-world financial systems and may reveal the financial activities of users.
To address this potential risk, all user identities of the Mercari and JPMC data have been encrypted by the corresponding owner companies prior to our use.
We are not able to share the Mercari and JPMC datasets but we are releasing the Venmo network.  Although Venmo transaction information is publicly available to anyone by the default user settings, we anonymized the users' names and IDs to prevent potential leakages of user information.

\section{Related Work}

We first briefly summarize the previous work on financial networks and fraud detection. Then we outline the related work on temporal network motifs.

\subsection{Financial Networks and Fraud Detection}

Advances in machine learning and data mining have introduced various computational methods for detecting financial fraud. When labeled data is available, classifiers are trained on features like transaction amount, user segmentation, and text~\cite{west2016intelligent,abdallah2016fraud,sadgali2019performance}. However, such features can be easily manipulated by fraudsters.

To address this, researchers have turned to network-based methods. Savage et al. used anomaly detection to identify money laundering through community structures~\cite{savage2017detection}. Van Vlasselaer et al. integrated intrinsic and network features like degrees and neighborhood similarity~\cite{van2017gotcha}. Li et al. developed metrics for detecting high-volume money flows~\cite{li2020flowscope}. Kodate et al. applied random forests to local network properties for e-commerce fraud detection~\cite{kodate2020detecting}. However, most methods ignore event chronology, which can be vital for identifying fraud.

Graph embedding methods also offer promise~\cite{LINE,node2vec}. Liu et al. applied GNNs to heterogeneous graphs of accounts and devices~\cite{liu2018heterogeneous}, while Yu et al. used random walks and auto-encoders to detect anomalies in dynamic graphs~\cite{yu2018netwalk}. These methods, though powerful, are computationally intensive and often lack explainability. Nonetheless, we include LINE~\cite{LINE} and node2vec~\cite{node2vec} in our experiments.

Beyond fraud, network analysis aids in studying financial systems. Zhang et al. analyzed Venmo's transaction graphs, identifying distinct patterns in user-to-user and user-to-vendor communities~\cite{zhang2017cold}. They compared graph metrics like degree, assortativity, and reciprocity but did not use transaction timestamps. They inferred vendor roles based on user behavior with strangers, without verifying vendor status. In contrast, our work predicts friendship ties using only transaction data, leverages temporal motifs to detect complex vendor patterns, and validates findings using vendor labels.

\subsection{Temporal Network Motifs}

Motifs have proven valuable in various domains involving temporal networks. Extending static motifs~\cite{Milo02,Przulj07}, several works adapted motifs for temporal contexts using snapshots. Jin et al. introduced trend motifs based on dynamic node weights, applying them to financial and protein networks~\cite{jin2007trend}. Chechik et al. used activity motifs---ordered combinations of chains, forks, and joins---to study gene interactions in yeast~\cite{chechik2008activity}. Zhao et al. developed communication motifs to analyze synchronous/asynchronous information flow in CDR and Facebook networks~\cite{zhao2010communication}. Bajardi et al. modeled cause-effect chains in cattle trade movements~\cite{bajardi2011dynamical}, and Zhang et al. analyzed diverse human interactions through extensive motif-driven studies~\cite{zhang2015human}. Kosyfaki et al. proposed flow motifs for edge-weighted temporal networks such as bitcoin, transit, and Facebook data~\cite{Kos19}.

The first temporal motif model for event streams was proposed by Kovanen et al., introducing temporal adjacency~\cite{K11}. Song et al. framed event pattern matching for streaming data, allowing partially ordered motifs~\cite{S14}. Hulovatyy et al. extended graphlet-based motifs with relaxed constraints on and support for event durations~\cite{H15}. Paranjape et al. presented a practical model bounded by a time window~\cite{P17}. Liu et al. surveyed and unified these models by combining temporal adjacency with time window constraints, and explored correlations between consecutive events~\cite{liu2021temporal}.

\section{Background}\label{sec:bg}
We explore temporal networks that we represent as $G = (V, E)$, where $V$ is the set of nodes, and $E$ is the set of time-stamped events.
Each event $e_i \in E$ is a 3-tuple $(u_i, v_i, t_i)$, which represents a directed relation from the source node $u_i$ to the target node $v_i$ that occurs at time $t_i$.
The set $E$ is a time-ordered list of $|E|$ events such that $t_1 \le t_2 \le t_3 \le \cdots \le t_{|E|}$.
The set of neighbors of a node $u$ in $G = (V, E)$ is defined to be $\Gamma(u)=\{v~|~(u, v, t_i) \in E~or~(v, u, t_j) \in E\}$ for some $t_i$ and $t_j$, i.e., we consider all the incoming and outgoing neighbors.
The one-hop egocentric network of a node $u$ is defined as $G_u = (V_u, E_u)$ where $V_u = \{u\} \cup \Gamma(u)$ and $E_u$ is the set of edges $(w, x)$ such that $w, x \in V_u$.

A temporal network can be projected to a static network by discarding the timestamps of the events. 
Formally, $\bar{G} = (V, \bar{E})$ is a (directed) static projection of $G = (V, E)$ iff $(u, v) \in \bar{E}$, then $(u, v, t) \in E$ for some $t$.
Here we distinguish edges and events, where the edge $\bar{e} = (u, v)$ is the static projection of an event $e = (u, v, t)$.
There can possibly be multiple events occurring on the same edge at different times.

\subsubsection{Temporal Motifs}
Given a temporal network $G = (V, E)$ and an inter-event time threshold $\Delta_C$, we define a $k$-node $l$-event temporal motif ($k \geq 2$ and $l \geq 2$), $M = (V', E')$, as a temporal subgraph in $G$ such that

\begin{itemize}
\item $|V'| = k$, $|E'| = l$, $V' \subseteq V$, and $E' \subseteq E$.
\item $E' = \{ (u'_1, v'_1, t'_1), \ldots, (u'_l, v'_l, t'_l) \}$ for $t'_1 \le t'_2 \le \cdots \le t'_l$
\item For any pair of consecutive events $(u'_i, v'_i, t'_i)$ and $(u'_{i+1}, v'_{i+1}, t'_{i+1}) \in E'$ such that \{$u'_i, v'_i\} \cap \{u'_{i+1}, v'_{i+1}\} \neq \emptyset$, it holds true that $t'_{i+1} - t'_{i} \leq \Delta_C$.
\item For any pair of events $(u',v',t') \in E'$ and $(u'', v'', t'') \in E'$, $t' \neq t''$.
\end{itemize}

\begin{figure}[!b]
\centering
\includegraphics[width=0.8\linewidth]{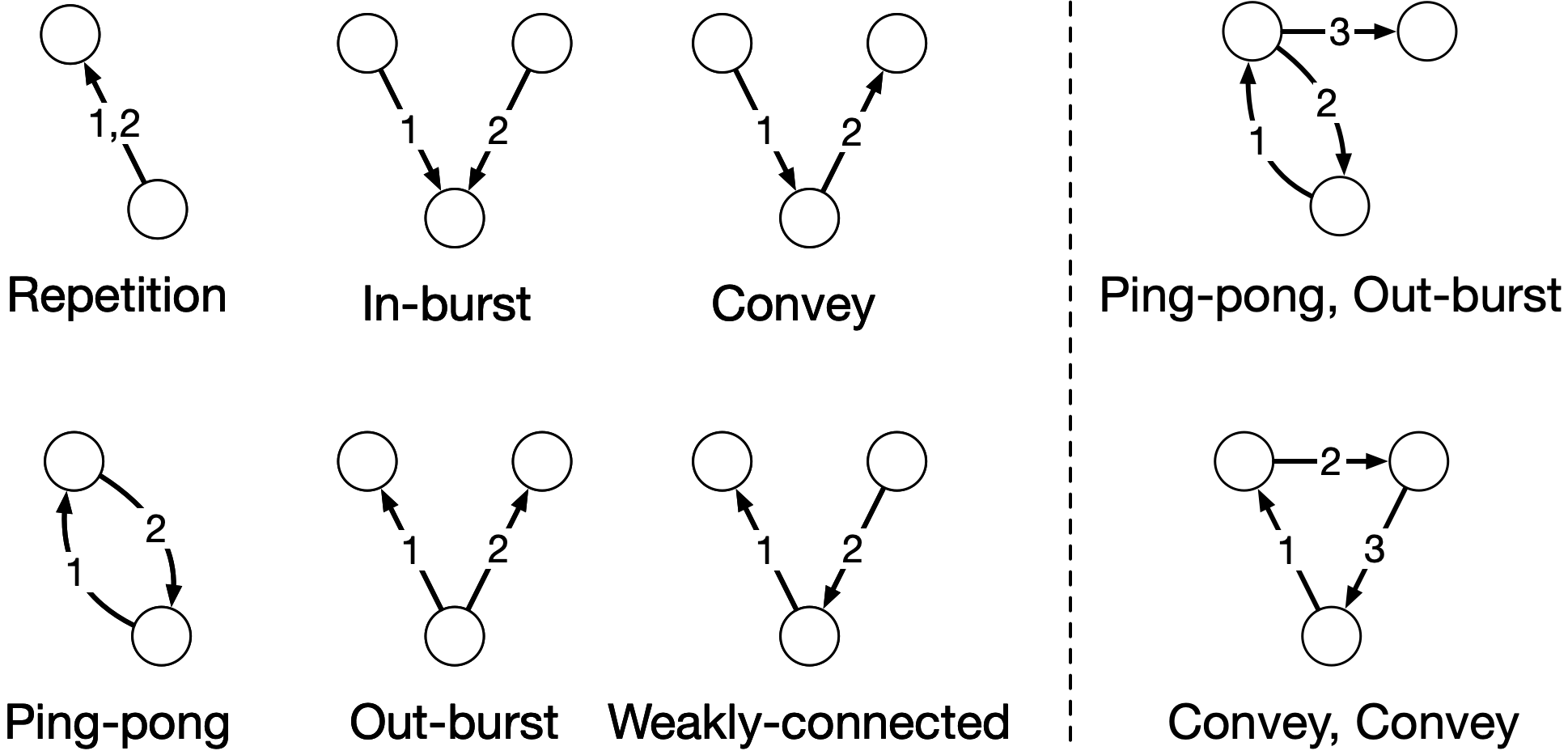}
\caption{\small [Left] All 2-event motifs, which are six types of consecutive event pairs. 
[Right] Two examples of 3-event motifs denoted by the sequence of event pairs.
}
\vspace{-4ex}
\label{fig:notation}
\end{figure}

Using a larger $\Delta_C$ threshold allows us to discover more temporal motifs. However, a large $\Delta_C$ has less power to control the relevance between consecutive events in the motif, especially for temporal networks with short timespans or inter-event times. In addition, increasing the $\Delta_C$ value will also exponentially increase the computation cost.
To address this trade-off, we choose $\Delta_C$ based on two characteristics of the temporal networks: the mean inter-event time and the connectivity rate of the consecutive events. For a temporal network with $|E|$ events, the mean inter-event time ($\delta$) is the average arrival time of consecutive events:
\vspace{-1ex}
$$
\delta = \frac{1}{|E|-1} \sum_{i=1}^{|E|} (t_{i+1} - t_i)
$$
We define the connectivity rate of a temporal network ($\gamma$) as the ratio of the number of consecutive events pairs $(u_i, v_i, t_i)$ and $(u_{i+1}, v_{i+1}, t_{i+1})$, such that $\{u_i, v_i\} \cap \{u_{i+1}, v_{i+1}\} \neq \emptyset$, to the total number of consecutive event pairs in the temporal network (i.e., $N-1$). 
We set $\Delta_C = \delta / \gamma$ so that $\frac{\Delta_C}{\delta} \times \gamma = 1$. In other words, within the time interval of $\Delta_C$, we are expected to find a pair of events that are consecutive and share a node.
For convenience, we round the calculated $\Delta_C$ value up to an hour in our experiments.

The 2-event motif is the simplest temporal motif by our definition, which is a pair of consecutive and connected events.
\cref{fig:notation} shows all types of 2-event motifs, named as repetitions, ping-pong, in-burst, out-burst, convey, or weakly-connected event pairs in~\cite{Liu22}.
A motif with $l$ events can be described as a sequence of $l-1$ event pairs.
\cref{fig:notation} gives an example of a 3-node 3-event motif which can be represented as a sequence of ping-pong (first and second events) and out-burst (second and third events), and a temporal cycle that is a sequence of two convey motifs.

\section{Fraud Detection}
In this section, we consider two datasets in which the fraud activity is defined differently: (1) financial transactions from an online marketplace, Mercari, where the fraudsters are the sellers of illegal items, and (2) a synthetic financial transaction network used by J.P. Morgan Chase, where the fraudsters are defined as the beneficiaries of the fraudulent transactions.
For both datasets, we utilize temporal motifs to detect fraudulent users. We extract temporal motif features from the egocentric networks and compare them against several baselines.

\begin{table*}[!t]
\centering
\begin{tabular}{|l||r|r|r|r|r|r|r|r|}
\hline
\multicolumn{1}{|c||}{Network}           & \multicolumn{1}{c|}{Nodes}  & \multicolumn{1}{c|}{Edges}   & \multicolumn{1}{c|}{Events} & \multicolumn{1}{c|}{Fraudsters}   & \multicolumn{1}{c|}{Time span}    & \multicolumn{1}{c|}{$\delta$} & \multicolumn{1}{c|}{$\gamma$}      & \multicolumn{1}{c|}{$\Delta_C$}     \\ \hline
Mercari           & 741,879 & 876,413   & 1,000,000 & 22.76\%   & 3/22/18-3/30/20 & 0.05s & 72.6\% & 1hr \\
JPMC           & 18,361  & 17,442   & 58,263 & 9.05\%  & 2/2/21-10/2/22 & 900.51s & 8.0\% & 3hr \\ \hline
\end{tabular}
\vspace{2ex}
\caption{\small Properties of the Mercari and JPMC networks.}
\label{tab:data}
\vspace{-5ex}
\end{table*}

\subsection{Online Marketplace Transactions}\label{sec:mercari}

\noindent {\bf Data.} We build the consumer-to-consumer online marketplace network of Mercari, which is one of the largest e-commerce platforms in Japan\footnote{Mercari, Inc. approved the use of the data for this study.}. Each user is represented as a node in the network and an event $(u, v, t)$ denotes that user $u$ sells an item to user $v$ at time $t$.
\cref{tab:data} shows the statistics of the Mercari network. 
The original data contains 84,285,577 trading transactions among 2,248,209 users from 3/31/2018 to 3/30/2020. In this study we select one million most recent transactions which cover the trading activities among 741,879 users in 8 days. There are 61,718 connected components in this sample, and the largest (weakly) connected component contains 80.0\% of the nodes in the network.
It is illegal to sell certain items in the Mercari marketplace, such as weapons, medicine, and used underwear.
The sellers of the prohibited items are considered as fraudulent users.
Our goal is to identify the fraudulent users based on their trading interactions with their neighbors.
We do not have access to any other information about user features or item properties, such as monetary amount or whether an item is illegal or not.\\

\noindent {\bf Temporal motif features.} We utilize the temporal motif features to distinguish whether a user is fraudulent or not in the online marketplace. For each user $u$, we first construct its egocentric network $G_u = (V_u, E_u)$, where $V_u$ contains $u$ and its one-hop neighbors, and $E_u$ includes all the events among nodes in $V_u$. Then we create the temporal motif features from $G_u$. In particular, we create a feature vector from $d$ types of temporal motifs $X_u = [x_1, x_2, x_3, \dots, x_d]$ s.t.
\begin{equation}
x_i = \frac{|M^i_u|}{|M^i_u| + |M^i_{\neg u}|}
\label{eq:feature}.
\end{equation}
In \cref{eq:feature}, $|M^i_u|$ is the number of type $M^i$ motifs in $G_u$ that contain node $u$, and $|M^i_{\neg u}|$ is the number of type $M^i$ motifs in $G_u$ that do not contain the node $u$. 
We create the feature vectors for the Mercari users based on all the 42 types of 2-event and 3-event motifs, with at most 3 nodes, hence $d=42$.
\cref{eq:feature} allows us to identify users who show significantly different trading patterns ($M^i_u$) than their neighbors ($M^i_{\neg u}$). 
Another benefit of~\cref{eq:feature} is that it naturally standardizes the count of different types of motifs in a range between 0 and 1, preventing the commonly observed motif patterns from being over-amplified.
The mean inter-event time of Mercari network is 0.05 seconds and the connectivity rate of consecutive events is 72.6\% (see \cref{tab:data}), hence we set $\Delta_C = 1$ hour.
According to \cite{P17}, the time complexity of counting 3-event temporal motifs is the number of static triangles times the number of temporal edges. Given an input temporal graph with $|V|$ nodes, $|\bar{E}|$ static edges, and $|E|$ events, the average degree of a node is $2{|\bar{E}|}/{|V|}$, and the maximum number of static edges in a one-hop egocentric graph is $O({|\bar{E}|^2}/{|V|^2})$. Therefore, the number of static triangles in the egocentric network is $O(({|\bar{E}|^2}/{|V|^2})^{1.5})=O({|\bar{E}|^3}/{|V|^3})$. Since each edge has ${|E|}/{|\bar{E}|}$ events on average, the number of events in the egocentric network is $O(({|\bar{E}|^2}/{|V|^2}) \times ({|E|}/{|\bar{E}|})) = O(({|\bar{E}|\cdot |E|})/{|V|^2})$. We repeat this for each node's egocentric network, so the time complexity of building temporal motif features is $O(|V| \times ({|\bar{E}|^3}/{|V|^3}) \times ({|\bar{E}|\cdot |E|})/{|V|^2}) = O({(|\bar{E}|^4\cdot |E|)}/{|V|^4})$. Note that real-world networks are often sparse, hence $|\bar{E}| << |V|^2$, and also the financial transaction networks in our datasets have $|\bar{E}| \approx |V|$ as shown in \cref{tab:data}.\\

\begin{table*}[!t]
\renewcommand{\tabcolsep}{1ex}
\centering
\resizebox{\columnwidth}{!}{
\begin{tabular}{|c||c||c|c|c|c|c|c|c||c|c|c|c|c|c|c|}
\hline
\multirow{4}{*}{Method} & \multirow{4}{*}{Classifier}  & \multicolumn{7}{c||}{Mercari}  & \multicolumn{7}{c|}{JPMC}  \\ 
& & \multicolumn{3}{c|}{Fraud}  & \multicolumn{3}{c|}{Non-Fraud} & AUC- & \multicolumn{3}{c|}{Fraud}  & \multicolumn{3}{c|}{Non-Fraud} & AUC-  \\
& & Pre. & Rec. & F1 & Pre. & Rec. & F1 & ROC & Pre. & Rec. & F1 & Pre. & Rec. & F1 & ROC \\   \hline

\multirow{4}{*}{LINE}	&	LR	&	0.105	&	0.543	&	0.176	&	0.893	&	0.453	&	0.601	&	0.498	&	0.028	&	0.395	&	0.052	&	0.972	&	0.605	&	0.746	&	0.500	\\
	&	SVM	&	0.107	&	0.346	&	0.164	&	0.896	&	0.663	&	0.762	&	0.504	&	0.031	&	0.285	&	0.056	&	0.971	&	0.728	&	0.832	&	0.506	\\
	&	RF	&	0.000	&	0.000	&	0.000	&	0.894	&	1.000	&	0.944	&	0.500	&	0.000	&	0.000	&	0.000	&	0.970	&	1.000	&	0.985	&	0.500	\\
	&	XGBoost	&	0.000	&	0.000	&	0.000	&	0.895	&	1.000	&	0.944	&	0.500	&	0.000	&	0.000	&	0.000	&	0.974	&	1.000	&	0.987	&	0.500	\\ \hline
																															
\multirow{4}{*}{node2vec}	&	LR	&	0.141	&	0.583	&	0.227	&	0.922	&	0.582	&	0.713	&	0.582	&	0.128	&	0.681	&	0.216	&	0.989	&	0.860	&	0.920	&	0.770	\\
	&	SVM	&	0.220	&	0.502	&	0.306	&	0.931	&	0.791	&	0.855	&	0.646	&	0.094	&	0.644	&	0.165	&	0.989	&	0.837	&	0.907	&	0.740	\\
	&	RF	&	0.359	&	0.001	&	0.002	&	0.896	&	1.000	&	0.945	&	0.500	&	0.000	&	0.000	&	0.000	&	0.972	&	1.000	&	0.986	&	0.500	\\
	&	XGBoost	&	0.304	&	0.010	&	0.019	&	0.894	&	0.997	&	0.943	&	0.503	&	0.471	&	0.133	&	0.208	&	0.977	&	0.996	&	0.986	&	0.565	\\ \hline
																															
	&	LR	&	0.820	&	0.640	&	0.720	&	0.850	&	0.940	&	0.890	&	0.789	&	1.000	&	0.020	&	0.040	&	0.810	&	1.000	&	0.900	&	0.509	\\
Simple	&	SVM	&	0.670	&	0.220	&	0.330	&	0.730	&	0.950	&	0.830	&	0.586	&	0.000	&	0.000	&	0.000	&	0.810	&	1.000	&	0.890	&	0.500	\\
 graph	&	RF	&	0.820	&	0.810	&	0.810	&	0.910	&	0.920	&	0.920	&	0.863	&	0.360	&	0.180	&	0.240	&	0.820	&	0.920	&	0.870	&	0.552	\\
	&	XGBoost	&	0.820	&	0.810	&	0.820	&	0.920	&	0.920	&	0.920	&	0.868	&	0.380	&	0.190	&	0.250	&	0.830	&	0.930	&	0.870	&	0.558	\\ \hline
																															
	&	LR	&	0.940	&	0.710	&	0.810	&	0.880	&	0.980	&	0.930	&	0.844	&	1.000	&	0.650	&	0.790	&	0.920	&	1.000	&	0.960	&	0.825	\\
Temporal	&	SVM	&	0.940	&	0.760	&	0.840	&	0.900	&	0.980	&	0.940	&	0.871	&	1.000	&	0.640	&	0.780	&	0.920	&	1.000	&	0.960	&	0.820	\\
 motif	&	RF	&	0.940	&	0.810	&	0.870	&	0.920	&	0.980	&	0.950	&	0.892	&	1.000	&	0.650	&	0.790	&	0.920	&	1.000	&	0.960	&	0.825	\\
	&	XGBoost	&	0.950	&	0.800	&	0.870	&	0.920	&	0.980	&	0.950	&	0.890	&	1.000	&	0.650	&	0.700	&	0.920	&	1.000	&	0.960	&	0.825	\\ \hline
																															
Simple 	&	LR	&	0.960	&	0.880	&	0.920	&	0.950	&	0.980	&	0.960	&	0.930	&	1.000	&	0.650	&	0.790	&	0.920	&	1.000	&	0.960	&	0.826	\\
graph and	&	SVM	&	0.960	&	0.710	&	0.820	&	0.880	&	0.990	&	0.930	&	0.848	&	1.000	&	0.610	&	0.760	&	0.920	&	1.000	&	0.960	&	0.807	\\
temporal	&	RF	&	0.950	&	0.950	&	0.950	&	0.980	&	0.980	&	0.980	&	0.962	&	0.860	&	0.670	&	0.750	&	0.930	&	0.970	&	0.950	&	0.820	\\
 motif	&	XGBoost	&	0.960	&	0.950	&	0.950	&	0.980	&	0.980	&	0.980	&	0.964	&	0.860	&	0.660	&	0.750	&	0.920	&	0.980	&	0.950	&	0.818	\\ \hline
		
\end{tabular}
}
\caption{\small Classification results on the Mercari and JPMC datasets for logistic regression, support vector machine, random forest, and XGBoost on five different variants: (1) LINE embeddings, (2) node2vec embeddings, (3) simple graph features, (4) temporal motif features, and (5) both simple graph and temporal features. Precision, recall, and F1 scores are given for both the fraud and non-fraud classes.}
\label{tab:allresults}
\vspace{-6ex}
\end{table*}

\noindent {\bf Baselines.} We compare our method with three baselines: (1) simple graph features~\cite{kodate2020detecting}, (2) LINE embeddings~\cite{LINE}, and (3) node2vec embeddings~\cite{node2vec}, all computed on the static projection of the graph.
Following~\cite{kodate2020detecting}, we extract ego-centric features for each node $u$: degree $k_u$, event count $s_u$, event per edge $s_u / k_u$, edge and event direction ratios $k^{\mathrm{out}}_u / k_u$, $s^{\mathrm{out}}_u / s_u$, local clustering coefficient, and cycle probability ${CYP}_u$. These features are used for fraud classification. Triangle counting dominates the computation with $O(|E|^{1.5})$ complexity. We train LINE and node2vec embeddings on the static undirected graph. node2vec uses 128 dimensions, walk length 80, 10 walks, $p=q=1$; runtime is $O(P(|V|+|E|))$. LINE uses 128 dimensions, batch size 1024, 11 epochs, $K=5$ negative samples, and has complexity of $O(dK|E|)$. We use the GraphEmbedding implementation~\footnote{\url{https://github.com/shenweichen/GraphEmbedding}}.\\

\noindent {\bf Experimental setup.} We performed all the experiments in this work on a Linux operating system running on machines with Intel(R) Xeon(R) Gold 6130 CPU processor at 2.10 GHz with 128 GB memory.\\

\noindent {\bf Results.} We use four different classifiers---logistic regression, support vector machine, random forest, and XGBoost---on five different variants: (1) LINE embeddings, (2) node2vec embeddings, (3) simple graph features, (4) temporal motif features, and (5) both simple graph and temporal features.
In XGBoost classifier, we use max depth as 3, choose 100 estimators, and set learning rate to 0.01.
For logistic regression, we use \textit{L2} penalty.
For random forest we use 100 estimators.
All classifiers are implemented using scikit-learn\footnote{\url{https://scikit-learn.org/stable/index.html}} and all remaining parameters set as default in the corresponding implementations.
For each run we use 75\% of the data for training and the remaining 25\% for testing.
For each configuration, we repeat the experiments 100 times and report the average precision, recall, and F1 scores for fraud and non-fraud classes, along with the AUC-ROC scores.
\cref{tab:allresults} presents the results.

\begin{figure}[!b]
\vspace{-2ex}
\centering
\includegraphics[width=\linewidth]{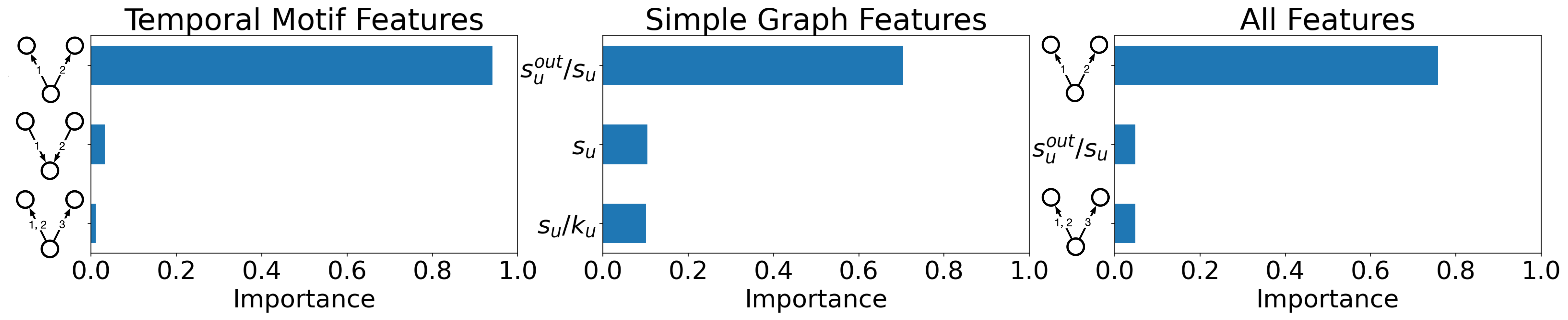}
\vspace{-4ex}
\caption{\small Importance of features in the online marketplace transactions in Mercari network. Each bar shows the importance of the feature determined by random forests classifiers. For each feature set, shown in each panel, we only show the top three features.}
\label{fig:importance-mercari}
\vspace{-4ex}
\end{figure}

Classifiers trained on temporal motif features consistently outperform those using simple graph features, LINE, or node2vec embeddings. They achieve up to 0.870 F1 for fraud, 0.950 for non-fraud, and 0.892 AUC. Temporal motifs also show robust performance across all classifiers. Adding simple graph features improves performance slightly, especially for XGBoost, which reaches 0.964 AUC.

\cref{fig:importance-mercari} shows the three most important features considered by the random forest classifier trained with temporal
motif features, simple graph features, and all features together. The number of out-going events divided by the total number of events $s^{\mathrm{out}}_u / s_u$ is the most important simple graph feature, while out-burst motifs are the most important temporal motif feature. The out-burst motifs still contribute most to the model’s classification power when considering all features together. In fact, on average, each non-fraudulent
user participated in 0.233 out-burst motifs whereas each fraudster participated in 1.256 out-burst motifs. This result suggests that the fraudulent merchants tend to sell their illegal products to multiple customers in a short time period.

\begin{table}[!t]
    \centering
    \footnotesize
    \begin{tabular}{|c||l|rr|}
    \hline
 &	&	Mercari  	&	JPMC	\\ \hline
\multirow{5}{*}{LINE}	&	{\it Embedding}	&	40943	&	22.00	\\
	&	LR	&	13.39	&	0.41	\\
	&	SVM	&	55778	&	40.92	\\
	&	RF	&	1165	&	10.77	\\
	&	XGBoost	&	316	&	8.04	\\ \hline
\multirow{5}{*}{node2vec}	& {\it Embedding}&	34699	&	1906	\\
	&	LR	&	6.85	&	0.37	\\
	&	SVM	&	60746	&	9.49	\\
	&	RF	&	778	&	19.66	\\
	&	XGBoost	&	294	&	13.18	\\ \hline
\multirow{4}{*}{Simple Graph}	&	LR	&	0.28	&	0.02	\\
	&	SVM	&	224	&	0.12	\\
	&	RF	&	4.16	&	0.03	\\
	&	XGBoost	&	1.47	&	0.32	\\ \hline
\multirow{5}{*}{Temporal Motif}	& {\it Motif counting}	&	36577	&	1195	\\ 
	&	LR	&	0.25	&	0.02	\\
	&	SVM	&	295	&	0.07	\\
	&	RF	&	5.79	&	0.21	\\
	&	XGBoost	&	5.58	&	0.34	\\ \hline
    \end{tabular}
    \caption{Classification and precomputation (embedding or motif counting) runtimes (seconds)}
    \label{tab:runtimes}  
    \vspace{-4ex}
\end{table}

We also investigate the runtime performance of all the methods. Table~\ref{tab:runtimes} presents all the classification runtimes as well as the precomputation runtimes for embedding-based methods (computing embeddings) and temporal motif features (computing motifs). The overhead of temporal motif computing for Mercari data is significant but not impractical when compared to the embedding times: motif counting is faster than LINE embedding computation by $\sim$50 mins and slower than node2vec embedding by $\sim$30 mins. Regarding the classification runtimes, SVM is the costliest one, taking $\sim$17 hours on node2vec embeddings. However, it takes only a few minutes on temporal motif features as the number of dimensions is much less when compared to the embedding methods (42 vs 128). Overall, temporal motifs is preferable to all the other methods.

\subsection{Synthetic Payment Transactions}

\noindent {\bf Data.} We consider a synthetic payment network generated by J.P. Morgan Chase (JPMC)\footnote{J.P. Morgan grants us the non-transferable limited license to use the data solely for this project.} using the characteristics of the real payment data \cite{assefa2020generating}. This is a high-fidelity synthetic data that preserves many unique features observed in a real transaction system.
The data includes 58,263 transactions among 18,361 users from 2021 to 2022 (see~\cref{tab:data}).
In addition to the transaction time and the anonymized IDs of the sender and receiver, each transaction also includes the monetary amount, the type of the payment method, and the users' countries of residence.
Payment fraud occurs when a fraudster deceives others to receive fraud payment to the fraudster's account. 
We mark a user as a fraudster if they receive at least one fraud payment.
Among 18,361 users in the JPMC data, 1,663 are labeled as fraudsters.
Similar to our approach for the Mercari data, we extract the temporal motif features and simple graph features from the one-hop egocentric network of each user, and also run LINE and node2vec embeddings. We train logistic regression, support vector machine, random forest, and XGBoost classifiers to predict if a user is a fraudster or not. 
The mean inter-event time of the JPMC network is 900.51 seconds, and the connectivity rate of consecutive events is 8.0\% (see \cref{tab:data}). Hence, we set $\Delta_C = 900.51/0.08 \approx 3$ hours when counting the 2-event and 3-event temporal motifs.\\

\begin{figure}[!t]
\includegraphics[width=\linewidth]{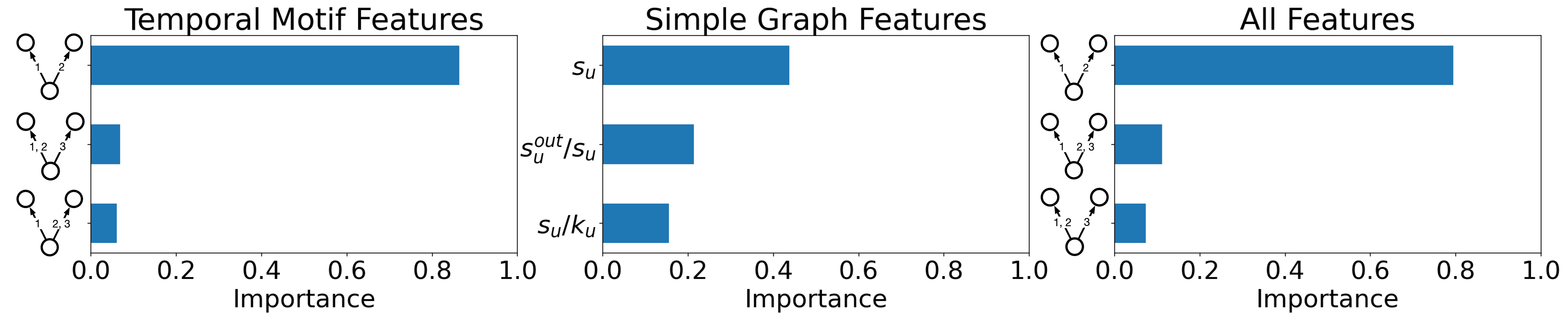}
\caption{\small Importance of the features in the synthetic payment transactions in JPMC network. Each bar shows the importance of the feature determined by random forests classifiers. For each feature set, shown in each panel, we only show the top three features.}
\vspace{-1ex}
\label{fig:importance-JPMC}
\end{figure}

\noindent {\bf Results.} Temporal motif features significantly improve fraud detection on the JPMC dataset (\cref{tab:allresults}), achieving up to 0.790 F1 for fraud and 0.960 for non-fraud, with 0.825 AUC. Simple graph features perform poorly---SVM fails entirely, and LR performs weakly. Embedding methods underperform, especially for fraud detection. Unlike in Mercari, adding simple features to motifs does not help, likely due to JPMC's smaller size (58K vs. 1M events).

\cref{fig:importance-JPMC} shows the three most important features considered by the random forest classifier trained on 
temporal motif features, simple graph features, and all features together. The out-burst motif is the most important 
temporal motif feature to support the decision of the fraud detection models. All simple graph features appear to 
have similar influences, while the number of events for each node ($s_u$) and the
ratio of out transactions ($s^{\mathrm{out}}_u / s_u$) are more important than the
other features. Among all temporal motif features and simple graph features, the out-burst motifs contribute most to predicting the fraudsters in the payment networks.

Regarding the runtime performance, the overhead of temporal motif counting is 40\% less than node2vec embedding, which suggests that the superior classification performance of temporal motif features is also practical. Also, the classification runtimes are less than a second for all temporal motif trainings whereas it takes up to 19 seconds to train node2vec embeddings.

\subsubsection{Monetary amounts in transactions}
\begin{figure}[!b]
\centering
\begin{subfigure}{0.49\linewidth}
\includegraphics[width=\linewidth]{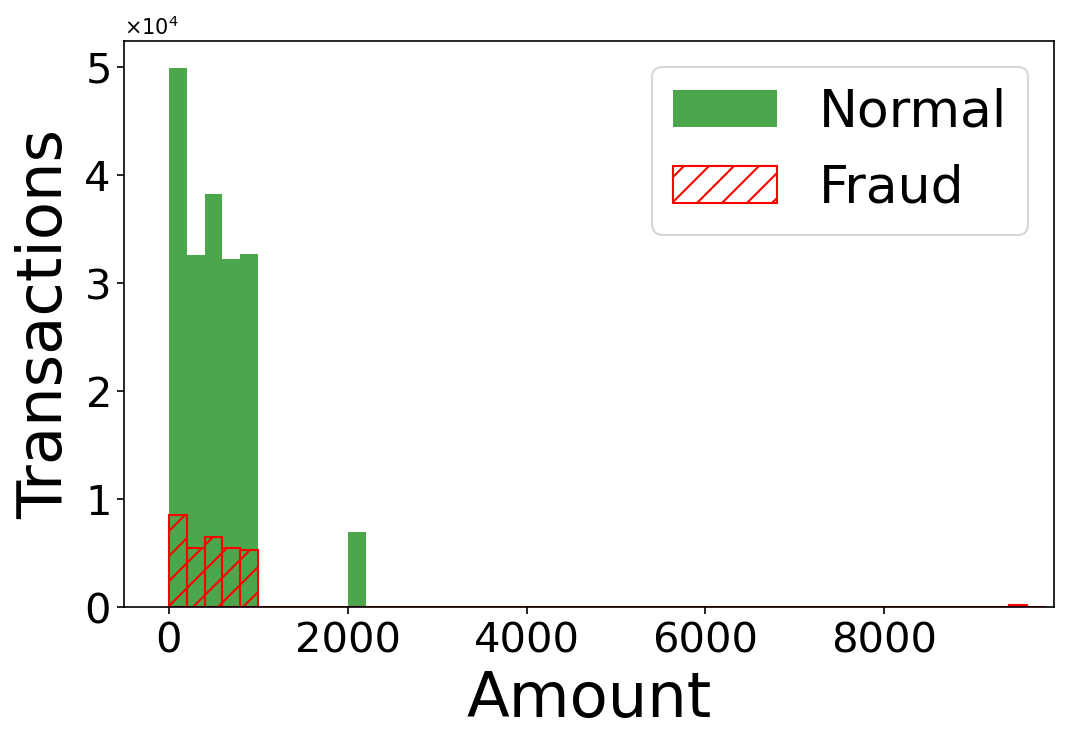}
\caption{\small Transactions from users.}
\end{subfigure}
\begin{subfigure}{0.49\linewidth}
\includegraphics[width=\linewidth]{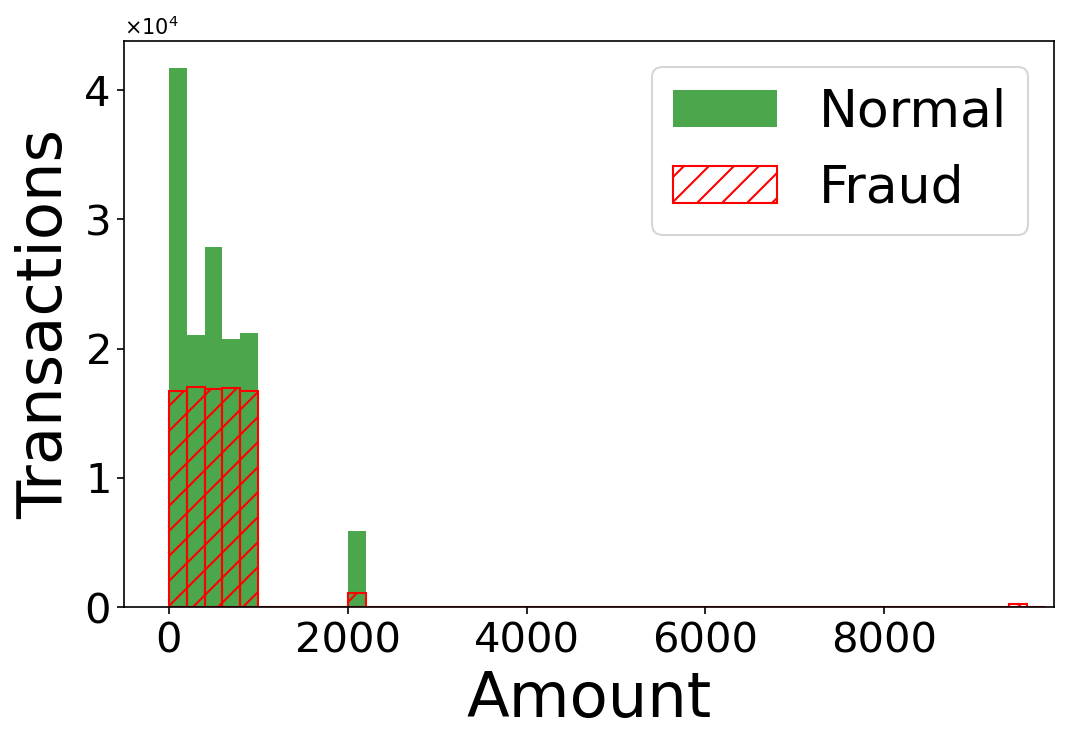}
\caption{\small Transactions to users.}
\end{subfigure}
\caption{\small The distribution of transaction amounts.}
\label{fig:amount}
\vspace{-3ex}
\end{figure}

As the JPMC data includes the monetary amounts of the transactions, we investigate the possibility of utilizing such information for fraud detection. We first check the monetary values in the transactions from normal/fraud users and to normal/fraud users.  
\cref{fig:amount} shows the histograms.
We observe similar amounts between transactions from and to normal users. There are also several transactions with large amount ($>\$9000$) that are sent to the fraudsters.

As the transaction amounts do not imply fraud activity on their own, we hypothesize that the temporal motifs that include fraudsters are more likely to have a different deviation in monetary amounts than the motifs that do not include fraudsters.
We use the coefficient of variation (CV), i.e., the standard deviation divided by mean, to measure the difference between the amounts of transactions in the motifs. Each type of motif $M^i$ is further categorized into three motifs based on its $CV$ value: $M^i_s$ if $CV \in [0, 0.5)$, $M^i_m$ if $CV \in [0.5, 1)$, and $M^i_l$ by $CV \in [1, \infty)$.
We use \cref{eq:feature} to create features for each $M^i$, thus defining $3*d$ number of motifs, and then train the classification models on the $3*d$-dimensional feature vectors to identify fraudsters. 

\cref{tab:amount} shows the results of the fraud detection trained on temporal motifs distinguished by the $CV$ of transaction amounts. Compared to the results in \cref{tab:allresults}, we observe a slight improvement in recall, AUC, and F1 scores.

\begin{table}[!t]
\small
\centering
\begin{tabular}{|l||r|r|r|r|}
\hline
Models           &   Precision   & Recall  & F1 Score & AUC-ROC               \\ \hline 
LR           &  1.0   & 0.690   & 0.817 & 0.845 \\ 
SVM           &  0.907   & 0.693    & 0.786 & 0.843  \\ 
RF 	   &  1.0   & 0.693    & 0.819 & 0.847       \\ 
XGBoost 	   &  1.0   & 0.693    & 0.819 & 0.847       \\ \hline      
\end{tabular}
\caption{\small Classification results on JPMC network with the monetary amount of transactions. Precision, recall, and F1 scores are given for the fraud class.}
\label{tab:amount}
\end{table}

\begin{figure}[!t]
\centering
\begin{subfigure}{0.49\linewidth}
\includegraphics[width=\linewidth]{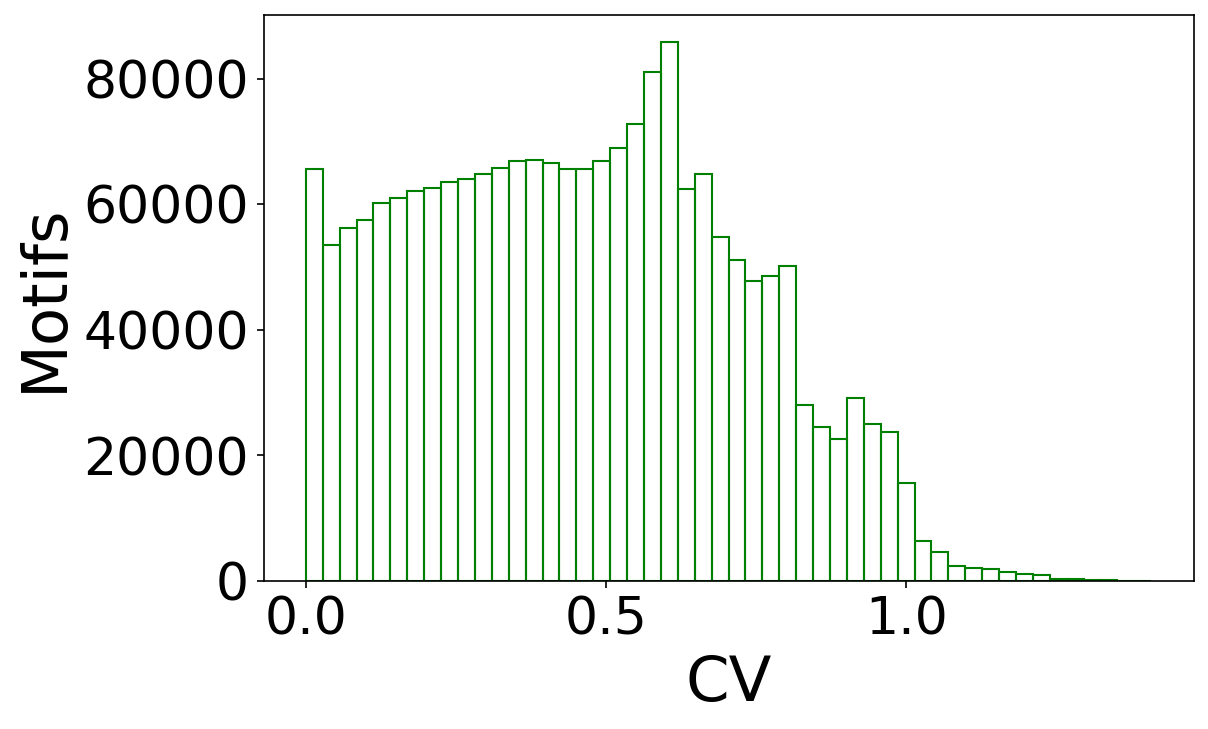}
\caption{Normal motifs.}
\label{fig:ping-pong}
\end{subfigure}
\begin{subfigure}{0.49\linewidth}
\includegraphics[width=\linewidth]{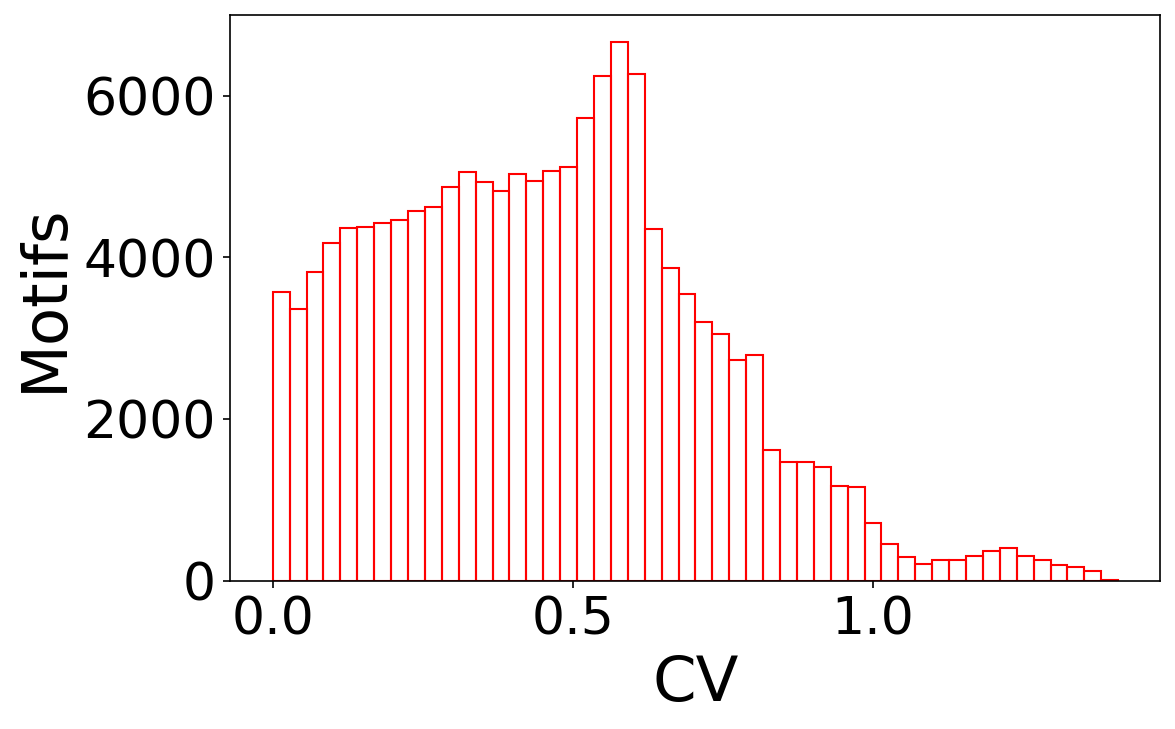}
\caption{Fraud motifs.}
\label{fig:in-burst}
\end{subfigure}
\vspace{-1ex}
\caption{\small The coefficient of variation (CV) of the transaction amounts in the normal and fraud temporal motifs in JPMC synthetic payment network.}
\label{fig:rsd}
\vspace{-2ex}
\end{figure}

\section{Interplay Between Transaction \& Friendship Networks}

Social and financial interactions are often intertwined---friendship may influence transactions and vice versa. Financial institutions leverage various networks, including social ones, to enhance fraud detection~\cite{fico}. In this section, we analyze Venmo data to explore the link between transaction and friendship networks using temporal motifs. We construct both networks via Venmo's public API and use motif features to (1) predict friendships based on transactions and (2) identify vendor users by combining transaction and friendship data. We also examine rare temporal motifs and associated payment notes for behavioral insights.

\subsection{Data}

Venmo, a leading digital payment platform, processes over a billion USD in monthly transactions. Users can publicly share transactions and form friend connections. We selected 600 highly active users from a public dataset of seven million Venmo transactions~\cite{venmo7m} and collected their full transaction histories using the Venmo API. This yielded 131,206 transactions among 19,141 users between 5/21/2015 and 2/9/2021. Each transaction includes sender and receiver IDs, timestamp, and a required message (though not the transaction amount).

\begin{table*}[!t]
\centering
\begin{tabular}{|l||r|r|r|r|r|r|r|}
\hline
\multicolumn{1}{|c||}{Network}          & \multicolumn{1}{c|}{Nodes}  & \multicolumn{1}{c|}{Edges}   & \multicolumn{1}{c|}{Events}  & \multicolumn{1}{c|}{Time span}    & \multicolumn{1}{c|}{$\delta$} & \multicolumn{1}{c|}{$\gamma$}     & \multicolumn{1}{c|}{$\Delta_C$}     \\ \hline
Venmo Transactions & 19,141  & 18,559   & 131,206  & 5/21/15-2/9/21 & 1721.61s  & 17.8\% & 24hr \\
Venmo Friendships  & 19,141 & 62,965   & N/A     & N/A & N/A  & N/A & N/A \\ \hline                    
\end{tabular}
\caption{\small Properties of the Venmo transaction and friendship networks.}
\label{tab:data-1}
\end{table*}

\cref{tab:data-1} presents summary statistics for the Venmo transaction and friendship networks. Friendship data is treated as static and undirected, as the API only indicates current friend status. The transaction network contains 227 connected components, with the largest covering 81.6\% of nodes. In contrast, the friendship network is more fragmented, with 1,271 components and a largest component covering 24.8\% of nodes.

\subsection{Supervised Friendship Discovery}

We use temporal motifs to predict whether two Venmo users are friends based on their transactions. For each node pair $(u, v)$, we count the number of 2-event and 3-event motifs that include both users. Only pairs with at least a transaction or friendship relation are included, totaling 52,021 pairs---86\% of which are friends.
The average inter-event time is 1,721.6 seconds, and 17.8\% of consecutive events are temporally connected. Given the five-year timespan, we set the motif time window $\Delta_C = 1$ day, slightly larger than $\delta / \gamma$, to capture meaningful patterns.\\

\begin{figure}[!t]
\centering
\begin{subfigure}{0.49\linewidth}
\includegraphics[width=\linewidth]{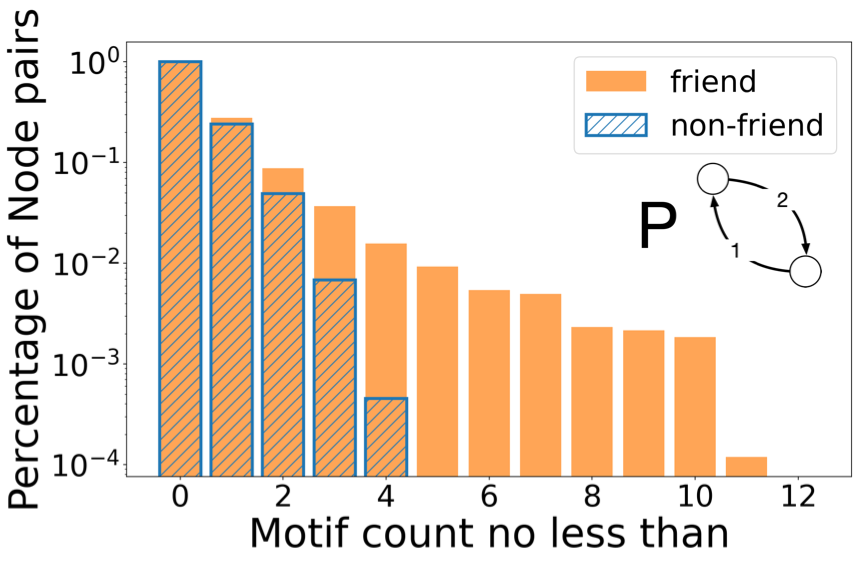}
\caption{\small Ping-pong}
\label{fig:ping-pong}
\end{subfigure}
\begin{subfigure}{0.49\linewidth}
\includegraphics[width=\linewidth]{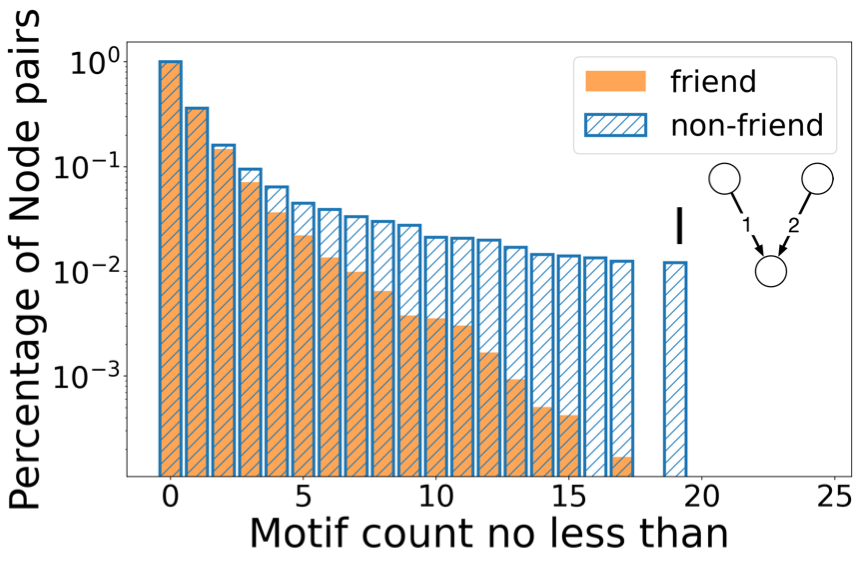}
\caption{\small In-burst}
\label{fig:in-burst}
\end{subfigure}
\vspace{-1ex}
\caption{\small The percentage of node pairs with motif count no less than the given amount. The horizontal axis represents the threshold amount, and the vertical axis shows the percentage of node pairs with the motif count no less than the threshold shown on the horizontal axis. No non-friend node pair has more than five ping-pong motifs, while half of the node pairs between friends are involved in more than five ping-pong motifs. On the other hand, we observe more than half of the non-friend node pairs are involved in more than 15 in-burst motifs.}
\label{fig:perc}
\vspace{-3ex}
\end{figure}

\noindent {\bf Temporal motif distributions.} We first examine if the motif counts between friend and non-friend node pairs show different patterns.
For each pair of nodes that are involved in an interaction (transaction or friendship), we count all the 2-event and 3-event temporal motifs and then investigate how the friend node pairs compare against non-friends.
We observe that motifs containing repetition, ping-pong, convey, and weakly-connected (remember~\cref{fig:notation}) are more dominant between friends, while motifs containing in-bursts and out-bursts occur more often between non-friend pairs.
\cref{fig:perc} shows the survival probability of the count of ping-pong and in-burst motifs for friend and non-friend node pairs. 
We do not observe any non-friend node pair that has more than 5 ping-pong motifs, while nearly 50\% of the friend node pairs are contained in at least 5 ping-pong motifs. On the other hand, in-burst motifs are more common in the non-friend node pairs than the between friend node pairs.\\

\noindent {\bf Features and baselines.} Based on our observations above, we extract features from the Venmo transaction network to predict the friendship relationship between users.
For each pair of nodes $u$ and $v$, we create a feature vector $X_{u, v} = [M^1_{u, v}, M^2_{u, v}, \dots, M^d_{u, v}]$ where $M^i_{u, v}$ is the number of $M^i$ motifs containing both $u$ and $v$.
We use all 2- and 3-event temporal motifs ($d=42$).
We train the logistic regression classification model with the temporal motif features to predict the friendship between two users.
We also consider two baseline models by applying the logistic regression model on the Jaccard (JC) coefficient~\cite{jac} and Adamic-Adar (AA) index~\cite{adamic2003friends}, which are commonly used heuristic scores for link prediction tasks.
For a pair of nodes $u$ and $v$, the Jaccard coefficient is
$J(u, v) = \frac{|\Gamma(u) \cap \Gamma(v)|}{|\Gamma(u) \cup \Gamma(v)|}$
and the Adamic-Adar index is measured as $A(u, v) = \sum_{w \in \Gamma(u) \cap \Gamma(v)} \frac{1}{\log |\Gamma(w)|}$,
where we remind that $\Gamma(u)$ is the set of neighbors of $u$. 
Note that these two scores do not consider temporality and work on the undirected static projection of the graph.
For classification, we use 75\% of the data to train and the rest to test.\\

\noindent {\bf Results.} Using temporal motif features achieves the best performance, with 0.86 precision, 1.0 recall, and 0.92 F1 scores. All three models are able to capture all the friend node pairs (with 1.0 recall scores), whereas temporal motifs yields less false positives than Jaccard and Adamic-Adar indices.

\subsection{Unsupervised Vendor Discovery}
As described in \cite{zhang2017cold}, Venmo users can be considered as customers or vendors, and friend transactions often show clear person-to-person patterns and stranger transactions usually indicate vendor-customer relations. 
Identification of vendors, who do not declare themselves as such, can be an important application for more accurate tax verification purposes as the current practice simply relies on a threshold on the number of transactions or the total amount received~\cite{venmotax}.
Here we utilize temporal motifs to identify vendor users.
We categorize Venmo transactions into two: 106,315 transactions between friends (TF) and 24,891 transactions between non-friend users (TN). Our goal is to develop an unsupervised approach to identify vendors in the Venmo network.\\

\begin{figure}[t!]
\centering
\begin{minipage}{0.5\linewidth}
  \centering
  \includegraphics[width=0.95\linewidth]{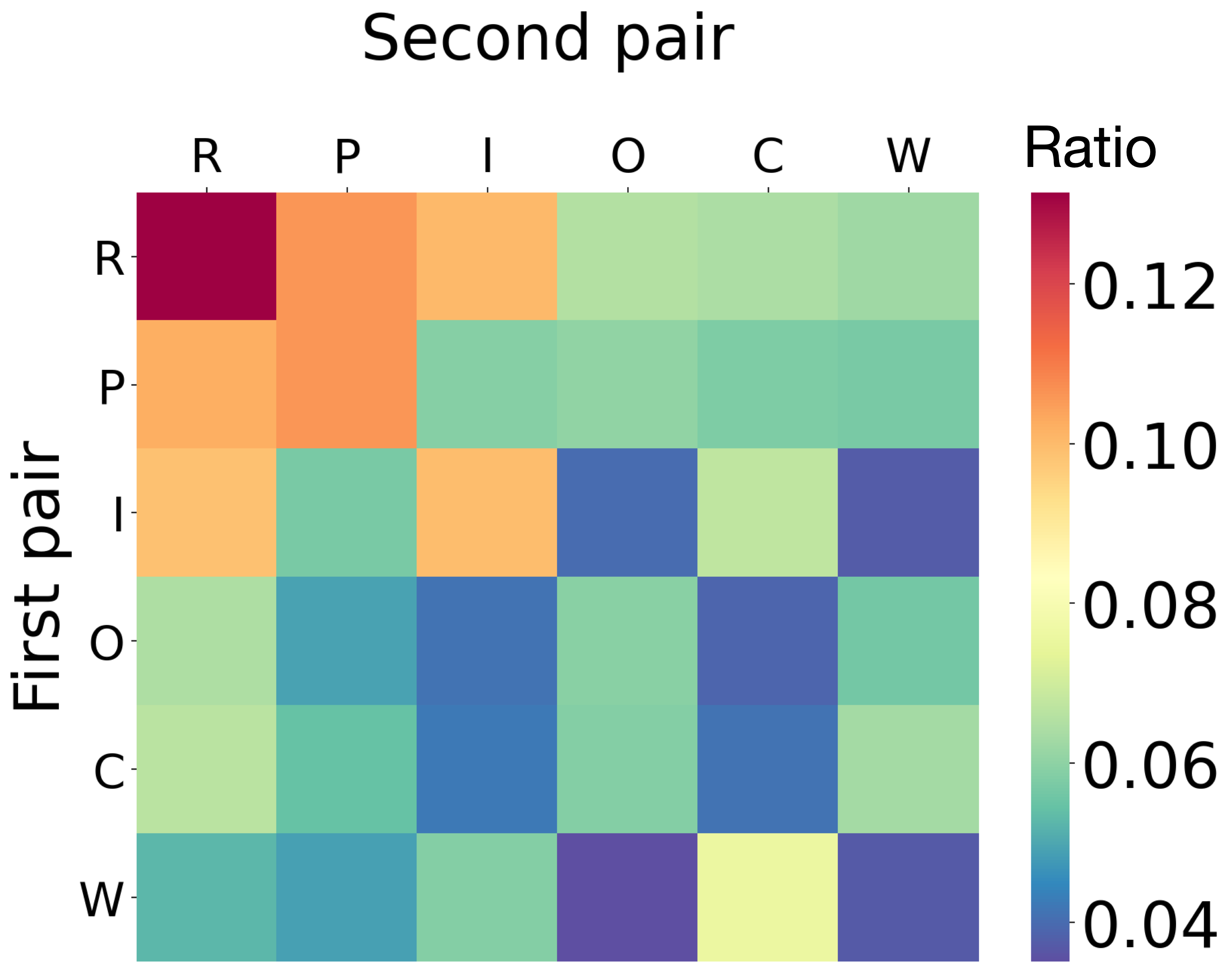}
\end{minipage}
\hfill
\begin{minipage}{0.45\linewidth}
  \small
  \vspace{4ex}
  \captionof{figure}{\small
    Ratios of the number of TN motifs to the number of TF motifs.  Each block represents a type of three-event motif; the row shows the first pair of events (first and second) and the column shows the second pair (second and third).  The ratios are color-coded. R, P, I, O, C, and W are repetition, ping-pong, in-burst, out-burst, convey, and weakly-connected, respectively.
  }
  \label{fig:heatmap}
\end{minipage}
\vspace{-4ex}
\end{figure}

\noindent {\bf Comparing TF and TN motifs.} We first compare the temporal motif counts in the TF and TN graphs.
Hereafter we refer the motifs in the TF and TN graphs as TF motifs and TN motifs, respectively.
For each type of 3-event motif, we compute the counts in both graphs by using $\Delta_C = 1$ day. We display the ratio of the count of the TN motifs to that of the TF motifs in~\cref{fig:heatmap}.
We observe that motifs that consist of repetitions, ping-pongs, and in-bursts are relatively more frequent in the TF graph than the TN graph. These motifs might correspond to real-world events in which a consumer purchases multiple times from the same vendor (repetition), a consumer asks for a refund from the vendor (ping-pong), or the vendor is paid by multiple consumers (in-burst). On the other hand, triangle motifs are more frequent in the TF graph than the TN graph, where three users are friends and make transactions with each other.\\

\begin{figure}[!b]
\centering
\begin{subfigure}{0.45\linewidth}
\includegraphics[width=\linewidth]{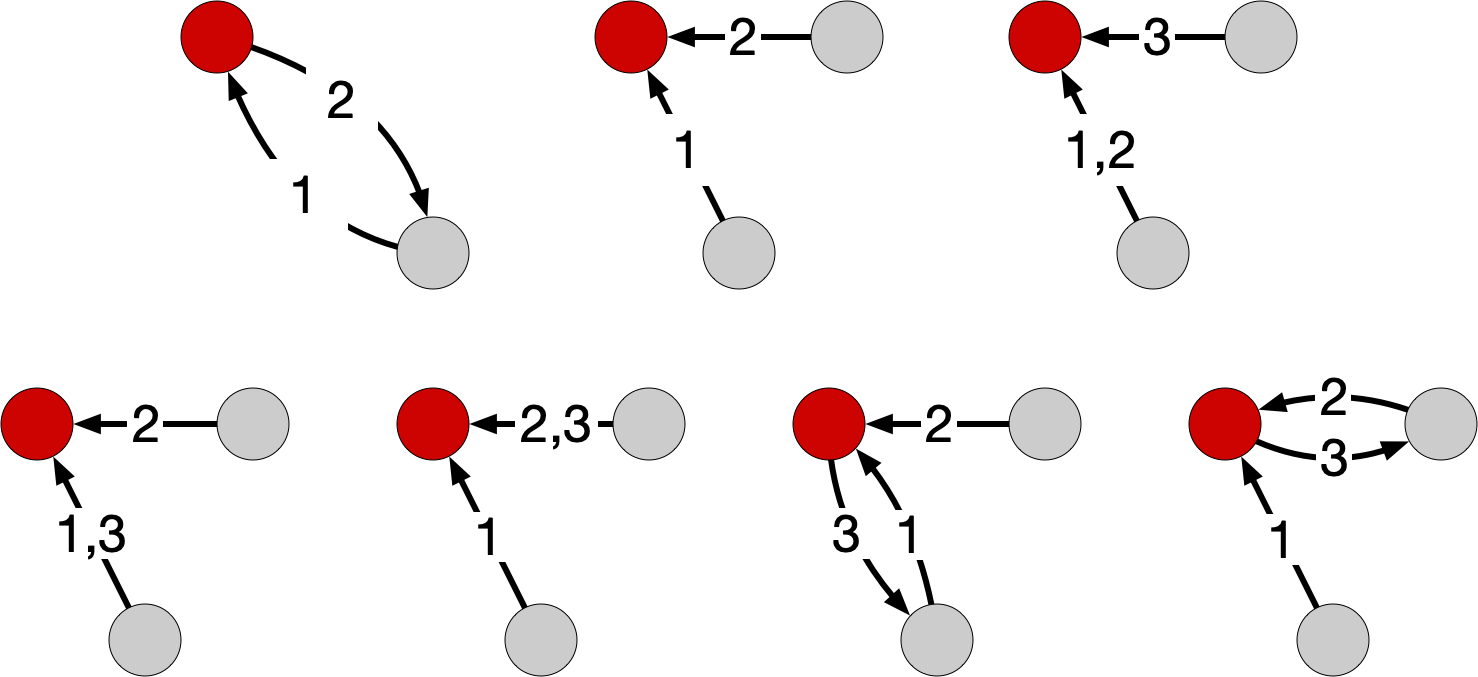}
\caption{\small Positive patterns}
\label{fig:positive}
\end{subfigure}
\hspace{3ex}
\begin{subfigure}{0.45\linewidth}
\includegraphics[width=\linewidth]{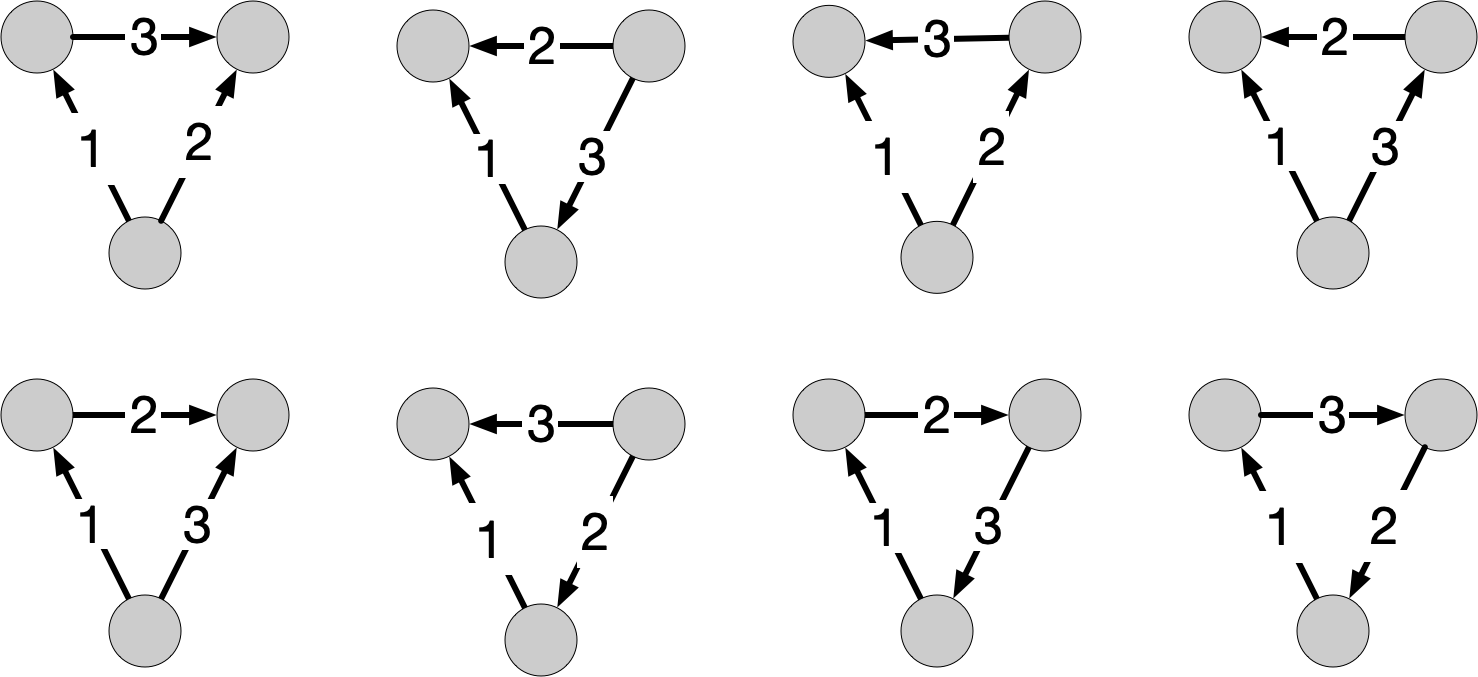}
\caption{\small Negative patterns}
\label{fig:negative}
\end{subfigure}
\caption{\small Vendor motif patterns. We select seven positive patterns in which the target node (red) is likely to be a vendor user (\cref{fig:positive}), and eight negative patterns that are unlikely to contain vendor users (\cref{fig:negative}).}
\vspace{-5ex}
\end{figure}

\noindent {\bf Vendor score.} We propose an unsupervised method to identify vendors in Venmo. For each user $u$, we compute a heuristic vendor score: $vs(u) = \log (|M^+_{TN}| + 1) - \log (|M^-_{TF}| + 1)$. 
Here, $M^+_{TN}$ denotes seven positive temporal motifs (shown in \cref{fig:positive}) likely associated with vendor behavior---such as in-burst motifs where a receiver rapidly transacts with multiple strangers, or ping-pong motifs suggesting refunds. We select three 3-event motifs where a target node receives payments from two strangers, and two 3-event motifs where a refund is implied. As shown in \cref{fig:heatmap}, repetition patterns are more common between strangers than friends. However, we exclude repetitive and ping-pong motifs from the score since they appear too frequently in both TF and TN networks and may overshadow more indicative patterns.
We define $M^-_{TF}$ as eight motifs (from \cref{fig:negative}) with the lowest TN-to-TF ratio---primarily triangles among friends---used to model anti-vendor behavior. The use of $\log$ helps manage motif count scale due to exponential growth with motif size.\\

\noindent {\bf Results.} To validate, we manually searched for profile information of the top 1,000 users by $vs(u)$. We identified 62 as vendors and 938 as non-vendors, while protecting personally identifiable data. Among the top 10 users, nine are verified vendors; among the top 20, 13 are verified.
~\cref{tab:vendor} gives the details.
The recall drops to 23\% and 6.1\% for top 100 and 1,000 users, respectively.
These results indicate that temporal motifs are a promising approach for vendor detection in transaction networks.

\begin{table}[!t]
\renewcommand{\tabcolsep}{1ex}
\small
\centering
\begin{tabular}{|l|c|l|}
\hline
$vs(u)$ & Vendor & Description\\ \hline
7.15  & Y  & a collegiate brand for students \\ 
6.03  & Y  & a college student organization \\ 
5.73  & Y  & a college student organization \\ 
5.66  & Y  & a travel agency \\ 
5.35  & Y  & a hotel       \\ 
5.31  & Y  & a bakery store       \\ 
5.27  & N  &  N/A        \\ 
5.25  & Y   & a geedunk bar    \\ 
5.21  & Y  & a business owner         \\ 
5.20  & Y & owner of an Amazon store \\ \hline
\end{tabular}
\caption{\small Ten Venmo users with the highest $vs(u)$ score.}
\label{tab:vendor}
\end{table}

\subsection{Curious Case of Temporal Cycles}
Temporal motifs can also be used to identify special activities in the Venmo network. For example, 3-node temporal cycles, $A \rightarrow B \rightarrow C \rightarrow A$ (convey-convey motifs), are the rarest type of 3-event motif in the Venmo transaction network.
Indeed such transactions are not expected to occur in a short time period; two transactions would be sufficient to balance all three accounts.
Among 637,439 3-event motifs in the transactions between friends, only 1.04\% are the 3-node temporal cycles.
Similarly, out of 49,682 3-event motifs among non-friends, only 0.55\% are temporal cycles.
We investigate the payment notes of the transactions which are involved in the 3-node temporal cycles between friends and identify 70 transactions with a note about ``poker".
For instance, \cref{tab:poker} shows six transactions among three individuals who are playing poker.
Venmo is indeed a popular medium (and legal in several states in the US) to exchange funds in online gambling platforms\footnote{\url{https://www.legalbettingonline.com/venmo/}}.
We believe that temporal cycles can be used to detect gambling activities and be helpful to law enforcement agencies in states and countries where online gambling is illegal.

\begin{table}[!t]
\small
\vspace{-4ex}
\renewcommand{\tabcolsep}{1ex}
\centering
\begin{tabular}{|c|c|c|l|}
\hline
Sender & Receiver & Time & Note\\ \hline
a  & b & 2020-03-23 1:34:10 & Poker \\ 
b  & c & 2020-03-23 3:27:41& Poker \\ 
c  & a & 2020-03-23 3:41:28 & poker\\ 
a  & b & 2020-04-11 1:49:50 & Poker \\ 
b  & c & 2020-04-11 2:49:25 & poker\\ 
b  & a & 2020-04-11 2:59:37 & Poker\\ \hline
\end{tabular}
\caption{\small Poker activities discovered by temporal cycles. The user IDs are anonymized. Note that ``Poker'' and ``poker'' are both original payment notes.}
\label{tab:poker}
\end{table}

\section{Conclusion}

This paper leverages temporal motifs to analyze financial networks and improve key applications. We extract motif-based features from users' egocentric networks and show they outperform static graph features and node embeddings in fraud detection on the Mercari and JPMC datasets. In Venmo, we use these features to predict friendships, identify vendors via an unsupervised approach, and uncover behaviors such as fund exchanges in social games. Our findings demonstrate the versatility and effectiveness of temporal motifs in understanding and modeling dynamic financial interactions.

\subsection{Limitations}

Our findings in this work are only limited to the three financial networks we have access to. Unfortunately, financial transaction networks are understudied, mainly due to their proprietary nature, as opposed to other types of real-world temporal networks in various domains.
We believe that cryptocurrency networks can offer some useful public data, although it is still challenging to verify the authenticity and find ground-truth information for fraudsters.
Another limitation of our work is that we have mainly focused on the temporal characteristics of the networks and compared against other network-based studies~\cite{kodate2020detecting}.
This does not mean the temporal motifs are the ultimate solution for fraud detection. We rather emphasize the importance of temporality in the analysis of financial networks.
We believe that temporal motifs can well be integrated to existing ML-based approaches for fraud detection on financial networks.
Regarding our specific method, we only extract the one-hop egocentric network to create temporal motif features. Extending the range of the egocentric network may yield useful temporal motifs which lie beyond the one-hop neighborhood. 

Another potential opportunity for future research would be improving the robustness of temporal motifs features. For instance, one can consider a null model for the transaction networks to normalize the motifs counts in the feature vectors.

\noindent \textbf{Acknowledgments.} This publication includes or references synthetic data provided by J.P. Morgan.
This research was supported by NSF-2107089 and NSF-2236789 awards, JP Morgan
Chase and Company Faculty Research Award, and used resources
of the CCR at the University at
Buffalo~\cite{CCR}.
We also thank Jingjing Chi and Yifan Wang for helping with the vendor labeling process.

\bibliographystyle{splncs04}
\bibliography{paper}

\end{document}